\documentstyle[epsf,12pt,graphicx]{article}
\begin{document}

\def\E{{\rm e}}
\newcommand{\be}{\begin{enumerate}}
\newcommand{\ee}{\end{enumerate}}
\newcommand{\beq}{\begin{equation}}
\newcommand{\eeq}{\end{equation}}
\newcommand{\uu}{\underline}

\centerline{\Large \bf Is there a universality}
\centerline{\Large \bf of the helix-coil transition}
\centerline{\Large \bf in protein  models?}
\vskip 1.1cm
\centerline{\bf Josh P. Kemp\dag , Ulrich H. E. Hansmann\ddag , 
                Zheng Yu Chen\dag}a
\vskip 0.3cm
\centerline{i\it \dag 
Dept.~of Physics, University of Waterloo, Waterloo, Ontario, N2L 3G1,
Canada}
\vskip 0.3cm
\centerline{\it \ddag Dept.~of Physics, Michigan Technological University,
Houghton, MI 49931-1291, USA}

\begin{abstract}
The similarity in the thermodynamic properties of
two completely different theoretical models for
the helix-coil transition is examined critically.
The first model is an all-atomic representation for a poly-alanine chain,
while
the second model is a minimal helix-forming model that contains no system
specifics.
Key characteristics of the helix-coil transition, in particular, the
effective
critical exponents of these two models agree with each other, within
a finite-size scaling analysis.

\end{abstract}
\centerline{Pacs: 87.15.He, 87.15-v, 64.70Cn, 02.50.Ng}

\baselineskip=0.8cm
The importance of understanding the statistical physics of the
protein-folding problem has been  stressed recently
\cite{DC,SHA4}.
For instance, it is now often assumed that the
energy landscape of a protein resembles a partially rough funnel.
Folding occurs by a
multi-pathway kinetics and the particulars of the folding funnel determine
the transitions between the different thermodynamic states
\cite{DC,Bryngelson87}.
This ``new view'' \cite{DC} of folding was
derived from  studies of minimal protein
models which capture only a few, but probably dominant parameters (chain
connectivity, excluded volume, etc.) in real proteins.

An implicit yet fundamentally crucial assumption is that the basic mechanism
of structural transitions in biological molecules depends solely on gross
features of the energy function, not on their details, and that a law of
corresponding states can be used to explain  dynamics and structural
properties of real proteins from studies of related minimal models. This
assumption needs to be proven.  An even stronger notion in statistical
physics is the universality hypothesis for critical phenomena.  The critical
exponents are identical for different theoretical models and realistic
systems belonging to the same universality class. Many theoretical concepts
in protein folding, such as coil-helix  or  coil-globular transitions
 involve phase transition or phase
transition-like behavior. Thus, one wonders if physical measurements between
two model systems for the same transition would have any
 ``universal'' properties.

The purpose of  this article is to examine these questions for the
helix-coil transition in homopolymers of amino acids \cite{Jeff,HO98c}.
Traditionaly, the coil-helix transition is
described by theories such as the Zimm-Bragg model \cite{ZB}
in which the homopolymers are regarded as one dimensional systems with only
local interactions; as such
a true thermodynamic phase transition is impossible.  However, recently there
have been
\cite{Jeff,HO98c} indications that the coil-helix transition near the
transition temperature displays phase-transition like behavior.
We use here finite-size scaling analysis, a common tool in statistical
physics, to examine
the question of universality of the helix-coil transition in two
completely different, illuminating models. On one hand, we have a
detailed, all-atomic representation of a homo poly-alanine chain
\cite{OH95b}. On the other hand, we have a simple coarse-grained model
describing the general features of helix-forming polymers \cite{Jeff}.
In this article, our interest lies in finding out how far the
similarity of the two models go.
 If the two models yield the same key
physical characteristics, then we at least have one concrete example of the
validity of the corresponding state principle or universality hypothesis in
biopolymer structures.

Poly-alanine is well-known to have high helix-propensities in proteins, as
demonstrated both experimentally and theoretically
\cite{HO98c,OH95b}.
It has been well tested and
generally believed that approximate force fields, such as ECEPP/2\cite{EC}
as implemented in the KONF90 program \cite{Konf}, give protein-structure
predictions to a surprisingly degree of faithfulness. As our first model, we
have ``synthesized'' poly-alanine with $N$ residues, in which the
peptide-bond dihedral angles were fixed at the value 180$^\circ$ for
simplicity. Since one can avoid the complications of electrostatic and
hydrogen-bond interactions of side chains with the solvent for alanine (a
non-polar amino acid), we follow earlier work \cite{OH95b} and neglect
explicit
solvent molecules in the current study.

Our second model is a minimalistic view of a helix forming polymer
\cite{Jeff} without atomic-level specifics. A wormlike chain is used to model
the backbone of the molecule, while a general directionalized interaction, in
terms of a simple square well form, is used to capture the essence of
hydrogen like bonding. The interaction energy between the residue labeled $i$
and $j$ is modeled by,
\begin{equation}
\label{minpot}
V_{ij}({\bf r}) = \left\{ \begin{array}{cl}
\infty & r < D \\
-v
 & D\leq r <\sigma \\
0 & \sigma\leq r
\end{array}
\right.
\end{equation}
where 
$v = \epsilon [{\bf \hat{u}}_i\cdot {\hat {\bf r}}_{ij}]^6
+ \epsilon [{\bf \hat{u}}_j\cdot {\hat {\bf r}}_{ij}]^6$, ${\bf
\hat{u}}_i=({\hat{
\bf r}}_{i+1,i})\times({\hat{\bf r}}_{i,i-1})$,
${\hat{ \bf r}}_{ij}$ is the unit vector between monomer $i$ and $j$, $D=
3/2a$ is the diameter of a monomer, $\sigma = \sqrt{45/8}a$ is the bonding
diameter,
and $a$ is the bond length while bond angle is fixed at $60^\circ$.

To obtain the thermodynamic properties, we have conducted multicanonical
Monte Carlo simulations for both models.  In the low-temperature region where
most of the structural changes occur, a typical thermal energy of the order
$k_BT$ is much less than a typical energy barrier that the polymer has to
overcome.  Hence, simple canonical Monte Carlo or molecular dynamics
simulations cannot sample statistically independent configurations separated
by energy barriers within a finite amount of available CPU time, and usually
give rise to bias statistics. One way to overcome this problem is the
application of {\it generalized ensemble} techniques \cite{Review}, such as
the {\it multicanonical algorithm} \cite{MU} used here, to the protein
folding problem, as has recently been utilized and
reported\cite{HO}.

In a multicanonical algorithm \cite{MU} conformations with energy $E$ are
assigned a weight $ w_{mu} (E)\propto 1/n(E)$, $n(E)$ being the density of
states.  A simulation with this weight generates a random walk in the energy
space; since a large range of energies are sampled,
one can use the re-weighting techniques \cite{FS} to calculate
thermodynamic quantities over a wide range of temperatures by
\begin{equation}
\left< {\cal{A}}\right>_T ~=~ \frac{\displaystyle{\int
dx~{\cal{A}}(x)~w_{mu}^{-1}(E(x))~e^{-\beta E(x)}}}
{\displaystyle{\int dx~w_{mu}^{-1}(E(x))~e^{-\beta E(x)}}}~,
\label{eqrw}
\end{equation}
where $x$ stands for configurations and $\beta$ is the inverse temperature.

In the case of poly-alanine chains, up to $N=30$ alanine residues were
considered. The multicanonical weight factors were determined by the
iterative procedure described in Refs.~\cite{MU} and we needed between
$4\times 10^5$ sweeps (for $N=10$)  and $5\times 10^5$ sweeps (for $N=30$)
for estimating
the weight factor approximately.  All thermodynamic quantities were measured
from a subsequent production run of $M$ Monte Carlo sweeps, where
$M$=$4\times 10^5$, $5\times 10^5$, $1\times 10^6$, and $3 \times 10^6$
sweeps for $N=10$, 15,20, and
30, respectively.  In the minimal model, chain lengths up to 39 monomers were
considered. In this model a single sweep involves a rotation of a group of
monomers via the pivot algorithm\cite{pivot}. For the weight factors the
similar
number of iterative procedure was used, and for the production run
$1\times 10^8$ sweeps was used in all cases.

We obtain the temperature dependence of the specific heat, $C(T)$,
by calculating
\begin{equation}
C(T)={\beta}^2 \ \frac{\left< E_{\rm tot}^2\right> - {\left<E_{\rm
tot}\right> }^2}{N}~,
\label{eqsh}
\end{equation}
where $E_{\rm tot}$ is the total energy of the system.  We also analyze the
order parameter $q$ which measures the helical content of a polymer
conformation and the susceptibility
\begin{equation}
\chi (T)  =  \frac{1}{N-2} (\langle q^2\rangle - \langle q\rangle^2
 )~.
\end{equation}
associated with $q$.  For poly-alanine $q$ is defined as
\begin{equation}
q = \tilde{n}_H
\end{equation}
where $\tilde{n}_H$ is the number of residues (other than the terminal ones)
 for which the dihedral angles ($\phi,\psi$) fall in the
range ($-70 \pm 20^{\circ},-37 \pm 20^{\circ}$). For our worm-like chain
model the order parameter $q$ is defined as
\begin{equation}
\label{wormorder}
q = \sum_{i=2}^{N-1} {\bf u}_i \cdot {\bf u}_{i+1}
\end{equation}
In both cases the first and last residues, which can move more freely,  are
not counted in the procedure.

From a finite-size scaling analysis of the heights and width of specific
heat
and susceptibility we can extract a set of effective critical exponents which
characterize the helix-coil transition in these two models \cite{fukugita}.
For instance,
with $C_{\rm MAX}$ defined to be the maximum peak in the specific heat,
 we have
 \begin{equation}
 C_{\rm MAX} \propto  N^{\displaystyle \frac{\alpha}{d\nu}}~.
 \label{alpha}
 \end{equation}
In a similar way, we find for the scaling of the maximum of the
susceptibility
 \begin {equation}
 \chi_{\rm MAX} \propto N^{\displaystyle \frac{\gamma}{d\nu}}~.
 \label{gamma}
 \end{equation}
For both quantities we  can also define the temperature gap
$\Gamma=T_2-T_1$ (where $T_1 < T_{\rm MAX} < T_2$) chosen such that $C(T_1)
=b C_{\rm MAX} = C(T_2)$, and $\chi (T_1)=b \chi (T_c) =\chi (T_2)$
 where $b$ is a fraction.
 The temperature gap obeys
 \begin{equation}
 \Gamma  = T_2 - T_1 \propto N^{\displaystyle -\frac{1}{d\nu}},
 \label{nu1}
 \end{equation}
as has been suggested in Ref.~\cite{fukugita} . The analysis should be
insensitive to the actual fraction, $b$, of $C_{\rm MAX}$ ($\chi_{\rm MAX}$)
considered
for defining $T_1$ and $T_2$
 which was verified from our numerical data
 fitting of poly-alanine chains.

The scaling exponents, $\alpha, \nu$, and $\gamma$, have their usual meaning
in critical phenomena; however, the above scaling relations also hold
formally
for the case of a first-order transition, with effective scaling
exponents $d\nu  = \alpha =  \gamma = 1$~\cite{fukugita,binder}.
Note that $d$ is the dimensionality of the system, and it always appears in
the combination $d\nu$. Without knowing further the effective dimensionality
of our systems, we use the combination $d\nu$ as a single parameter in the
fit.

It then becomes straightforward to use the above equation and the values
given in Table~\ref{tab1} to estimate the critical
exponents. We
obtain for poly-alanine from the scaling of the width of the specific heat
$1/d\nu=1.02(11)$ with a goodness of fit $(Q=0.9)$ (see Ref.~\cite{NR} for
the definition of $Q$), for chains of length $N=15$ to $N=30$. Inclusion of
$N=10$ leads to $1/d\nu=0.84(7)$, but with a less acceptable fit $(Q=0.1)$.
Similarly, we find from the scaling of the width of the susceptibility
$1/d\nu = 0.98(11)$ $(Q=0.5)$ for chains of length $N=15$ to $N=30$ and
$1/d\nu=0.81(7)$ $(Q=0.2)$ when the shortest chain $N=10$ is included in the
fit. Hence, we present as our final estimate for the correlation exponent of
poly-alanine $d\nu=1.00(9)$. This value is in good agreement with the
estimate $d\nu =0.93(5)$ obtained from the partition function zero analysis
in Ref.~\cite{AH99b}.

The results for the exponent $\alpha$ give $\alpha = 0.89(12)$ (Q=0.9) when
all chains
are considered, and $\alpha = 0.86(10)$ $(Q=0.9)$ when the shortest chain is
excluded from the fit. Analyzing the peak in the susceptibility we find
$\gamma = 1.06(14)$ $(Q=0.5)$ for chain lengths $N=15-30$ and $\gamma =
1.04(11)$ $(Q=0.5)$ for chain lengths $N=10-30$.
We summarize our final estimates for the critical exponents in Table
\ref{tab5}.  The scaling plot for the susceptibility
is shown in Fig.~1: curves for all lengths of
poly-alanine chains collapse on each other indicating the validity of finite
size scaling of our poly-alanine data.

The same procedure can be applied to analyze the data from the minimal model.
All calculation has been done with the omission of the
shortest chain. Using the widths of the specific heat a $b=80\%$ of the peak
height we obtain $1/d\nu~=~1.03(7)$, $(Q=0.2)$.  The
width of the
peak at half maximum is more unreliable in this case as the coil-helix
transition is complicated by the additional collapsing transition to a
globular state in the vicinity of the coil-helix transition\cite{Jeff}. This
exponent agrees with that calculated from the susceptibility widths,
$1/d\nu~=~0.89(9)$, $(Q=0.3)$.  Hence, our final estimate for this critical
exponent in
our second model is $d\nu = 0.96(8)$.  These values are in good agreement
with those of the poly-alanine model.

From the $C_{\rm MAX}$ data in Table~\ref{tab1} and using the above given
value for
the exponent $d\nu$ we find $\alpha~=~0.70(16)$ ($Q=0.3$) which is somewhat
smaller than that of the poly-alanine model. The susceptibility exponent as
calculated from the data in Table~\ref{tab1} yields a value of
$\gamma~=~1.3(2)$
($Q=0.5$), which agrees with the previous estimation within the error bar.
The scaling plot for the susceptibility is shown in
Fig.~2. While curves corresponding to large polymer
sizes collapse into the same curve, the $N=13$ case shows small disagreement,
indicating that the finite size scaling are valid only for longer chain
lengths in the minimal model.

Comparing the critical exponents of our two models as summarized in Table
\ref{tab5} we see that the estimates for the correlation exponent $d\nu$
agrees well for the two models.  Within the error bars, the estimates for the
susceptibility exponent $\gamma$ also agree.  The estimates for the specific
heat exponent $\alpha$ seem disagree within the error ranges. However, in
view of the fact that both analyses are based on small system size the true
error ranges could be actually larger than the ones quoted here.  Using these
rather crude results, we have already demonstrated a striking similarity in
finite-size scalings of the two model. Therefore, we can convincingly make
the conjecture that minimal model can be used to represent the structural
behavior of real helix-forming proteins.

Our analysis should tell us also whether the helix-coil
transition in our models is of first or second order.
In the former case we
would expect $d\nu = \alpha = \gamma = 1$
which seems barely supported by our
data due to the rather large error bars associated with the estimate of the
exponents.
We have further explored the nature of the transition from another
perspective, by considering the change in energy crossing a small temperature
gap (taken to be within 90\% of $C_{\rm MAX}$) from the original data,
\begin{equation}
\label{delE}
\Delta E = (E_{\rm tot}(T_2) - E_{\rm tot}(T_1))/N
\end{equation}
This value should approach either a finite value or zero as
$N^{-1}$ goes to zero. A finite value would indicate a first order transition
while a zero value a second order transition. In the case of a first order
transitions the intercept would indicate the latent heat.  Now, the
assumption is that this energy change scales linearly as $N^{-1}$ goes to
zero. Figure 3 shows a plot of the data from
both the atomic-level and minimal models, where nonzero intercepts can be
extrapolated at $N^{-1}$=0.
Hence, our results seem to indicate and  finite latent heat and a
first-order like helix-coil transition. However, we can not
exclude the possibility that the true asymptotic limit of
 $|E|$ is zero, and some of the results of Ref.~\cite{HO98c} point for
the case of poly-alanine rather towards a second-order transition.
Further simulations of larger chains seem to be necessary to determine the
order of the helix-coil transition without further doubts.

In summary, we conclude that in view of the similarity of the two models
examined here, a corresponding state principle can be established for the
coil-helix transition. Examining the finite size scaling analysis  allows
us to calculate estimators for critical exponents 
 in the two models which indicate ``universality'' of helix-coil transitions.

\noindent
{\bf Acknowledgments}:
Financial supports from Natural Science and Engineering Research Council of
Canada and the National Science Foundation (CHE-9981874)
are gratefully acknowledged.

\newpage
{\Large Figure Captions}\\
\begin{enumerate}
\item Scaling plot for the susceptibility $\chi(T)$
      as a function of temperature $T$, for poly-alanine
      molecules of chain lengths $N = 10, 15, 20$ and $30$.
\item Scaling plot of $\chi(T)$
      as a function of temperature $T$, for the minimum model
      of chain lengths $N = 13, 19, 26, 33$ and $39$.
\item Scaling of energy gap and transition width at 80\% and 90\%
 of $C_{MAX}$. Here we have used $\Delta E_{80\%}$($\bigtriangleup$ for
 all-atom model,$\Diamond$ for minimal model), $\Delta E_{90\%}$($\Box$ for
 all-atom model, $\bigcirc$ for minimal model).
\end{enumerate}

\newpage
\begin{table}
\caption{Shown are the location of
the specific heat maximum $T_{\rm MAX}$, the maximum of specific heat $C_{\rm
MAX}$, susceptibility $\chi_{\rm MAX}$, the width of the half peak in
specific
heat $\Gamma_{C}$, and width of the half peak of susceptibility
$\Gamma_{\chi}$ for various chain lengths.}
\begin{center}
\begin{tabular}{|c|c|c|c|c|c|}
\hline
$N$&$T_{\rm MAX}$& $C_{\rm MAX}$ & $\Gamma_{C}$ & $\chi_{\rm MAX} $ &
$\Gamma_{\chi}$\\
\hline
\multicolumn{6}{|c|}{All-Atomic Model}\\\hline
10& 427(7)& 8.9(3) & 160(7) & 0.49(2) & 140(7)\\
15& 492(5)& 12.3(4)& 119(5) &  0.72(3)& 110(5)\\
20& 508(5)& 16.0(8)& 88(5)  &  1.08(3)& 78(5)\\
30& 518(7)& 22.8(1.2)&58(4)&  1.50(8)& 56(3)\\
\hline
\multicolumn{6}{|c|}{Minimal Model}\\\hline
13& 1.25(1) & 1.088(2) & 1.22(2) & 0.232(2) & 2.20(2) \\
19& 1.17(1) & 1.424(5) & 1.12(2) & 0.353(3) & 0.81(2) \\
26& 1.16(1) & 1.789(8) & 0.89(2) & 0.553(8) & 0.57(2) \\
33& 1.13(1) & 2.08(1) & 0.73(2) & 0.78(1) & 0.45(2) \\
39& 1.12(1) & 2.27(2) & 0.61(2) & 0.96(2) & 0.41(2) \\\hline
\end{tabular}
\end{center}
\label{tab1}
\end{table}

\begin{table}
\caption{Summary of the critical exponents obtained for the two models.}
\begin{center}
\begin{tabular}{|c|c|c|}
\hline
   & All-atomic  &  Minimal \\
\hline
 $d\nu$   &  1.00(9)  &  0.96(8)\\
 $\alpha$ &  0.89(12) &  0.70(16)\\
 $\gamma$ &  1.06(14) &  1.3(2)\\\hline
\end{tabular}
\end{center}
\label{tab5}
\end{table}

\pagebreak
\begin{center}
\includegraphics[width=6.0in]{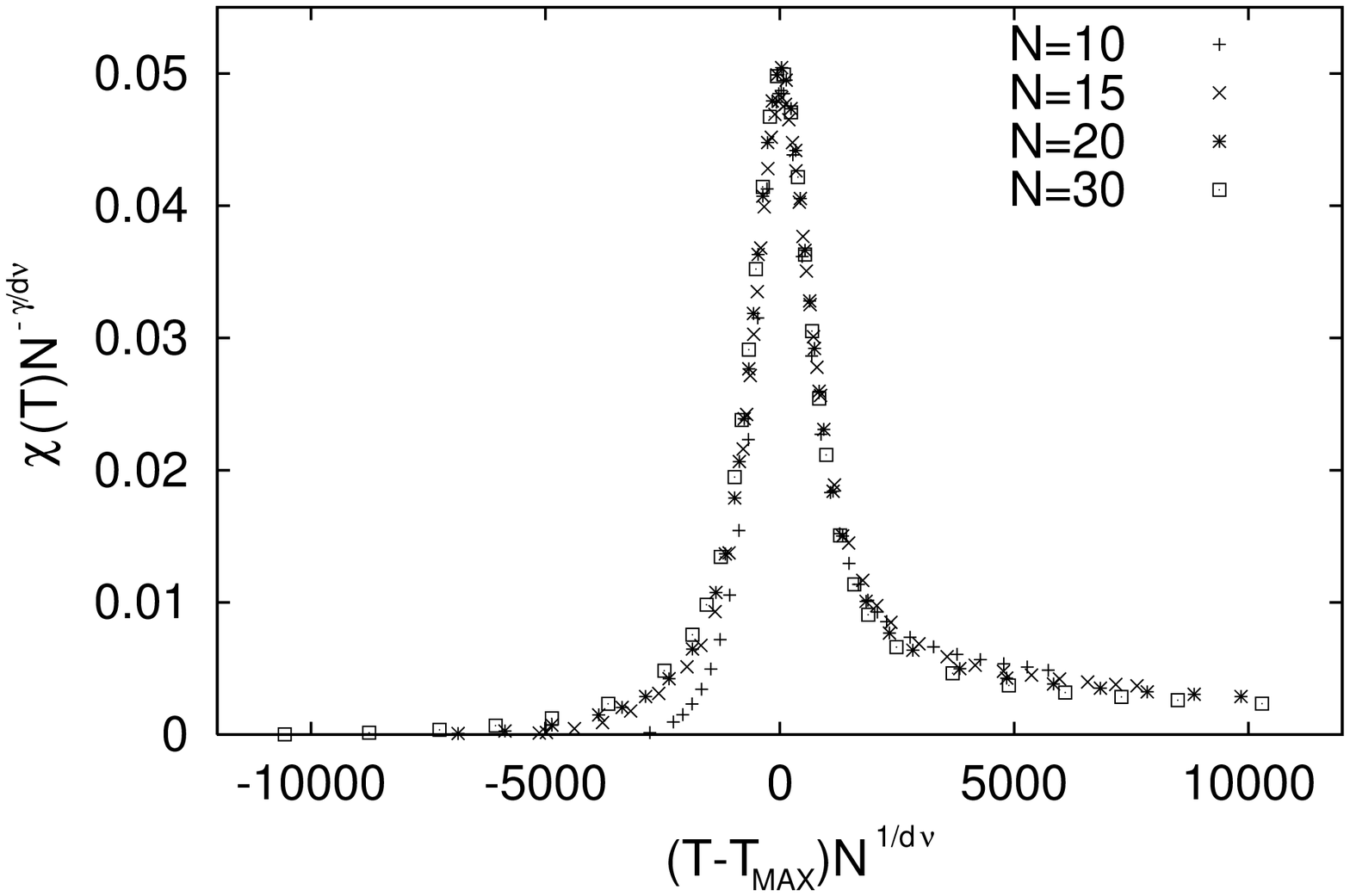}
\end{center}
\vspace{2in}
\begin{center}
{\Large Fig.~1  }
\end{center}

\pagebreak
\begin{center}
\includegraphics[width=6.0in]{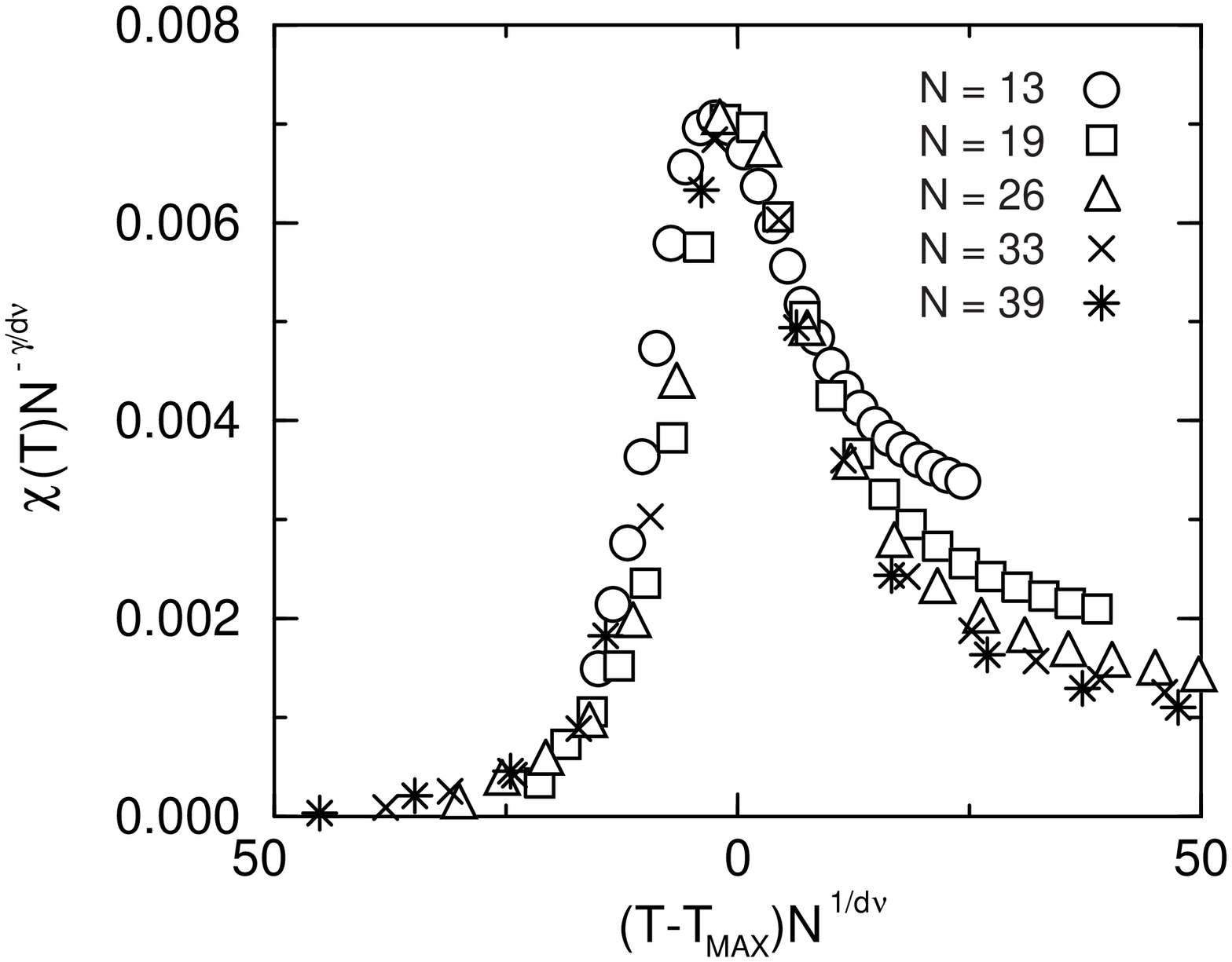}
\end{center}
\vspace{1.5in}
\begin{center}
{\Large Fig.~2 }
\end{center}

\pagebreak
\begin{center}
\includegraphics[width=6.0in]{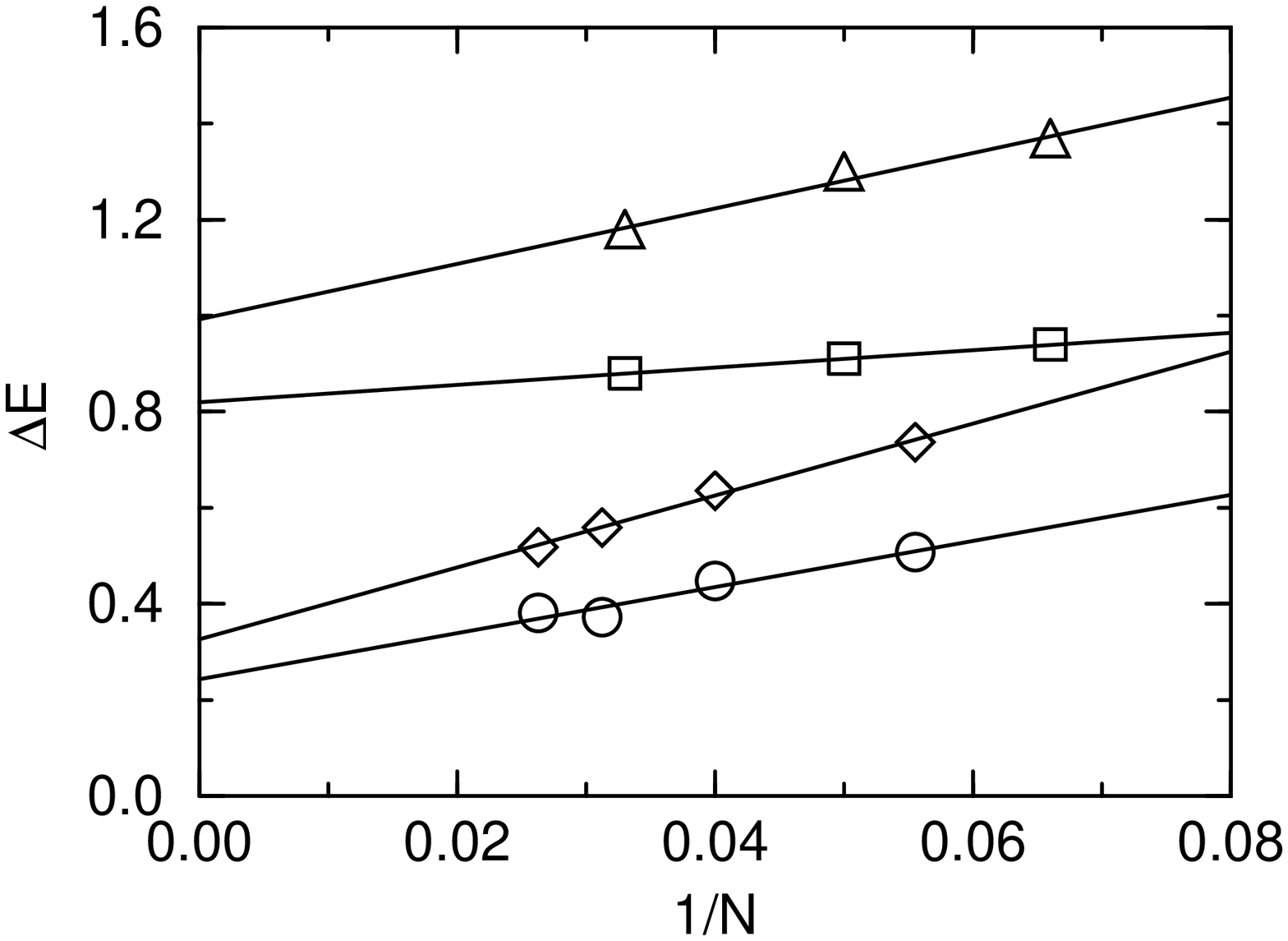}
\end{center}
\vspace{1in}
\begin{center}
{\Large Fig.~3  }
\end{center}

\end{document}